\documentstyle[amssymb,aps,12pt]{revtex}

\begin{document}
\author{David D. Reid\thanks{%
Email address: phy\_reid@online.emich.edu}}
\address{Department of Physics and Astronomy, Eastern Michigan University, Ypsilanti,%
\\
MI 48197\vspace{0.5in}}
\title{The Manifold Dimension of a Causal Set: tests in conformally flat spacetimes}
\maketitle

\begin{abstract}
This paper describes an approach that uses flat-spacetime estimators to
estimate the manifold dimension of causal sets that can be faithfully
embedded into curved spacetimes. The approach is invariant under
coarse-graining and can be implemented independently of any specific curved
spacetime. Results are given based on causal sets generated by random
sprinklings into conformally flat spacetimes in 2, 3, and 4 dimensions, as
well as one generated by a percolation dynamics.\vspace{12pt}\newline
PACS: 02.90+p, 04.60-m
\end{abstract}

\section{Introduction}

Since the time of Einstein, the prospect that spacetime might be discrete on
microscopic scales has been considered as one possible avenue to help solve
the problem of quantum gravity. The causal set program proposes one approach
to discrete quantum gravity [1-2]. A causal set is a set $C$ of elements $%
x_i\in C,$ and an order relation $\prec $, such that the set $C=\{x_i,\prec
\}$ obeys properties which make it a good discrete counterpart for continuum
spacetime. These properties are that (a) the set is transitive: $x_i\prec
x_j\prec x_k\Rightarrow x_i\prec x_k$; (b) it is noncircular, $x_i\prec x_j$
and $x_j\prec x_i\Rightarrow x_i=x_j$; (c) it is locally finite such that
the number of elements between any two ordered elements $x_i\prec x_j$ is
finite, i.e., $\left| [x_i,x_j]\right| <\infty $; and (d) it is reflexive, $%
x_i\prec x_i$ $\forall $ $x\in C$. The action of the order relation is to
mimic the causal ordering of events in macroscopic spacetime. Since all
events in spacetime are not causally related, then not all pairs of elements
in the set are ordered by the order relation. Hence a causal set is a
partially ordered set.

If the microscopic structure of spacetime is that of a causal set, then in
appropriate macroscopic limits, causal sets must be consistent with the
properties of general relativity which describes spacetime as a Lorentzian
manifold. Therefore, it must be established that causal sets can possess
manifold-like properties. A necessary (but not sufficient) requirement for a
causal set to be like a manifold is that it can be embedded into a manifold
uniformly with respect to the metric. Finding ways to embed a causal set has
proven to be very difficult thus far. However, the properties of a causal
set can be compared to the properties that a uniformly embedded causal set
are expected to have. The kinds of tests that can check for manifold-like
behavior generally require knowledge of the dimension of the manifold into
which the causal set might embed. In fact, consistency between different
ways to estimate the dimension of the manifold is itself a stringent test of
manifold-like behavior. It is worth noting that within the mathematics of
partial orders there are several types of dimensions. However, the
dimensions traditionally studied by mathematicians do not correspond to what
is meant here. Therefore, everywhere in this paper the phrase ``dimension of
a causal set'' refers to the {\it manifold dimension, }i.e.{\it ,} the
dimension of the Lorentzian manifold into which the causal set might be
uniformly embedded.

The most useful methods for estimating the dimension of a causal set are the
Myrheim-Meyer dimension [3-4] and the midpoint scaling dimension [5]. By
design, both of these methods work best in Minkowski space. The approach
used to derive the Myrheim-Meyer dimension has been extended to curved
spacetimes [4], but implementation of this more general {\it Hausdorff
dimension} is specific to the particular spacetime against which the causal
set is being checked. Since there are infinitely many curved spacetimes,
this method is less useful in more generic cases. Therefore, there is a
continuing need to find ways to estimate the dimension of a causal set for
curved spacetimes that (a) are independent of the specific properties of the
curved spacetime, (b) do not require very large causal sets to achieve
useful results, and (c) are invariant under coarse-graining of the causal
set. This last requirement is desired because, on the microscopic scale, the
causal sets that might describe quantum gravity will not display
manifold-like properties in the sense described above. Only in the
macroscopic limit, after an appropriate change-of-scale, do we expect to see
such properties; this change-of-scale is called coarse-graining.

In what follows, I first present the background theory and terminology
needed to understand the dimension estimation methods described in this
paper; then, the different approaches to dimension estimation are described.
The extent to which the methods work are illustrated using causal sets
generated by uniform sprinklings into flat and conformally flat spacetimes.
I then illustrate the methods using a causal set generated by a percolation
dynamics.

\section{Theory}

As alluded to previously, in the causal set program we are interested in
those causal sets that can be uniformly embedded into a manifold. An
embedding of a causal set is a mapping of the set onto points in a
Lorentzian manifold such that the lightcone structure of the manifold
preserves the ordering of the set. With high probability, an embedding will
be uniform if the mapping corresponds to selecting points in the manifold
via a Poisson process (as described below). Two important results for
understanding the dimension estimators to be discussed are (a) the
correspondence between the volume of a region in a manifold and the number
of causal set elements and, (b) the correspondence between geodesic length
and the number of links in a chain of causal set elements. These topics are
discussed in the next two subsections.

\subsection{Random Sprinklings}

One way to generate a causal set that can be uniformly embedded into a
manifold is to perform a random sprinkling of points in a manifold. If the
set $X$ consists of points $\{x_i\}$ randomly distributed (sprinkled) in a
manifold $M$ of finite volume $V_M$, we can define a discrete random
variable $\chi _A$ on a region $A$ of $M$ such that $\chi _A(x_i)=1$ if $%
x_i\in A$ and $0$ otherwise. In terms of the random variable $\chi _A$ we
can define another discrete random variable $N_n$ that counts the number of $%
x_i\in A$ up to a possible number $n$ equal to the size of $A$, where $%
N_n=\sum_{i=1}^{i=n}\chi _A(x_i)$. Random variables such as $N_n$ are
described by the binomial distribution which, for our case, can be written
as [4] 
\begin{equation}
F_k(N_n)=%
{n \choose k}%
\left( \frac{V_A}{V_M}\right) ^k\left( 1-\frac{V_A}{V_M}\right) ^{n-k},
\label{1}
\end{equation}
where $F_k$ is the probability of outcome $k$.

If we define the density of the sprinkled points as $\rho =n/V_M$, the
expectation value of $N_n$ in region $A$ is given by $\left\langle
N_n\right\rangle =\rho V_A$. To generalize this description to manifolds of
infinite volume, we take the limit of eq. (1) as $V_M\rightarrow \infty $
while holding the density of the sprinkling uniform, $\rho =const$. This
procedure is a standard approach for deriving the Poisson distribution [6] 
\begin{equation}
P_k(N_n)=\lim\limits_{%
{V_M\rightarrow \infty  \atop \rho =const}%
}F_k(N)=\frac{\rho V_A}{k!}e^{-\rho V_A},  \label{Poisson}
\end{equation}
where the equivalence $V_A/V_M=\rho V_A/n$ has been used. From this
distribution, we find that the average value of the number of points
sprinkled into regions of volume $V_A$ is given by 
\begin{equation}
\left\langle N_n\right\rangle _A=\rho V_A.  \label{num and vol}
\end{equation}
While it is customary to scale the sprinkling to unit density, $\rho =1$,
this scaling is not done in the present cases. Thus, we see that a random
sprinkling of points in a manifold at uniform density is described by a
Poisson distribution. Therefore, the interesting causal sets are from among
those that will admit an embedding consistent with a Poisson sprinkling into
a manifold (perhaps only after coarse-graining). Such an embedding is
referred to as a {\it faithful} {\it embedding}.

\subsection{Geodesic Length}

Recall that the length of the geodesic between two causally related events
corresponds to the longest proper time between those events. To see what the
most natural analog to geodesic length is for causal sets we must first
define a few terms. A {\it link,} $\preceq $, in a causal set is an
irreducible relation; so, $x_i\preceq x_k$ iff $\nexists $ $x_j\ni x_i\prec
x_j\prec x_k$. A {\it chain} in a causal set is a set of elements for which
each pair is related; for example, $x_a\prec x_b\prec \cdots \prec
x_{z-1}\prec x_z$ is a chain from $x_a$ to $x_z$. A {\it maximal chain} is a
chain consisting only of links, such as $x_a\preceq x_b\preceq \cdots
\preceq x_{z-1}\preceq x_z$.

As explained by Myrheim [3], the length of the longest maximal chain between
two related elements in a causal set is the most natural analog for the
geodesic length between two causally connected events in spacetime. (Myrheim
did not use the term ``causal set'' which was coined by Rafael Sorkin and
used in [1]). The length of a maximal chain is defined to be the number of
links in that chain. Brightwell and coworkers have proven that this
correspondence between the geodesic length in a Lorentzian manifold and the
number of links in the longest maximal chain is, in fact, valid in Minkowski
space [7]. Therefore, Myrheim's expectation that this correspondence should
be valid in the general case seems well founded. In this work, I shall
assume the validity of what I will refer to as the {\it Myrheim length
conjecture}:

\begin{quote}
Let $C=\{x_i,\prec \}$ be a causal set that can be faithfully embedded, with
density $\rho $, into a Lorentzian manifold $M$ by a map $g:C\rightarrow M$.
Then, in the limit $\rho \rightarrow \infty $, the expected length of the
longest maximal chain between any ordered pair $(x_i,x_j)\in C$ is directly
proportional to the geodesic length between their images $[g(x_i),g(x_j)]\in
M$.
\end{quote}

\section{Dimension Estimators}

A dimension estimator for a causal set is a method that only uses properties
of the set to determine the dimension of the manifold into which the causal
set might be faithfully embeddable. Ideally, we hope to have a scheme for
estimating the dimension of a causal set that (a) works well for curved
spacetime manifolds, (b) is invariant under coarse-grainings of the causal
set, and (c) does not require very large causal sets in order to see useful
results. As alluded to previously, one difficulty in finding a useful
dimension estimator for curved spacetimes is that implementation of the
estimators tend to depend on the properties of the particular spacetime
against which the causal set is being compared. This circumstance is
problematic for causal sets generated by a process that does not directly
suggest a class of candidate spacetimes.

However, one property that all physical spacetimes share is that locally,
they are approximately Minkowskian. From the standpoint of causal sets, this
implies that if a causal set $C$, of size $N$, is faithfully embeddable into
a $d$-dimensional curved manifold $M^d$, then there ought to be subsets $%
{\tt c}_i\subset C$, of size $n_i<N$ that are faithfully embeddable
(approximately) into $d$-dimensional Minkowski space ${\Bbb M}^d$. Studying
how these subsets behave under dimension estimators that work reliably for $%
{\Bbb M}^d$ should allow us to identify which, if any, $d$-dimensional
Minkowski space is most closely approximated by these subsets. I will refer
to dimensions found in the above manner as the {\it local Minkowski dimension%
} of the causal set. An approach similar to this was independently suggested
by Sorkin [8].

The dimension estimators that will be used to determine the local Minkowski
dimension in curved spacetimes are the Myrheim-Meyer dimension and the
midpoint-scaling dimension mentioned in the introduction and described
below. Both of these dimension estimators are defined in terms of causal set
intervals. A causal set interval between two related elements $I[y,z]$ is
the inclusive subset $I[y,z]=\{x_i|y\prec x_i\prec z\}$. Taking $\prec $ as
a causal order, $I[y,z]$ is the intersection of the future of $y$ with the
past of $z$.

The Myrheim-Meyer dimension is based on the fact that for a causal set
faithfully embeddable into an interval $I$ of ${\Bbb M}^d$, the expected
number of chains that consists of $k$ elements, $k$-chains ($S_k$), is given
by [4] 
\begin{equation}
<S_k>=\frac{\left( \rho V_I\right) ^k\Gamma \left( \delta \right) \Gamma
(2\delta )\Gamma \left( 2\delta +1\right) ^{k-1}}{2^{k-1}k\Gamma \left(
k\delta \right) \Gamma \left( (k+1)\delta \right) },  \label{4}
\end{equation}
where $\delta \equiv (d+1)/2$. The easiest chains to count are 2-chains
which count the relations between elements. Specializing to 2-chains, eq.
(4) becomes 
\begin{equation}
f(d)\equiv \frac{<S_2>}{<N>^2}=\frac{\Gamma \left( d+1\right) \Gamma \left(
d/2\right) }{4\Gamma \left( 3d/2\right) },  \label{ordering fraction}
\end{equation}
where I have used eq. (3) to relate number and volume. Therefore, for a
given causal set, we can divide the number of relations $S_2\approx <S_2>$
by the square of the number of elements $N\approx <N>$ to approximate the
value of $f(d)$ for the interval. This function is monotonically decreasing
with $d$ and can be numerically inverted to give a value for the dimension.

The midpoint-scaling dimension relies on the correspondence between number
and volume, and on the relationship between the volume of an interval in $%
{\Bbb M}^d$ and the length of the geodesic $\tau $ between its defining
events [5] 
\begin{equation}
V_I=\frac{\pi ^{(d-1)/2}}{2^{d-2}d(d-1)\Gamma \left[ (d-1)/2\right] }\tau ^d.
\label{6}
\end{equation}
An interval $I[y,z]$ of size $N$ can be divided into two sub-intervals $%
I_1[y,x]$ and $I_2[x,z]$ of sizes $N_1$ and $N_2$, respectively. Let $%
N_{small}$ be the smaller of $N_1$ and $N_2$, then the element $x$ is the
midpoint of $I$ when $N_{small}$ is as large as possible. This process
corresponds to a rescaling of lengths by a factor of $1/2$; therefore, in
the manifold $\tau /\tau _{small}=2$, which implies that $V/V_{small}=2^d$.
For the causal set interval, assuming the Myrheim length conjecture, this
translates to $N/N_{small}\approx 2^d$ so that 
\begin{equation}
d\approx \log _2\left( N/N_{small}\right)  \label{midpoint_d}
\end{equation}
estimates the dimension.

\section{Results}

The dimension estimators were applied to causal set intervals generated by
random sprinklings into flat and conformally flat spacetimes given by the
metric 
\begin{equation}
ds^2=\Omega ^2\eta _{\alpha \beta }dx^\alpha dx^\beta ,
\label{conformal metric}
\end{equation}
where $\Omega ^2$ is the conformal factor (a smooth, strictly positive
function of the spacetime coordinates) and $\eta _{\alpha \beta }$ is the
Minkowski tensor. The sprinklings were performed by two different methods.
The more efficient approach for sprinkling $N$ points into an interval $I$
of volume $V_I$ was to divide the interval into several little regions of
volume $v_i$. The number of points sprinkled into a region $n_i$ was
determined by the ratio $n_i/N=v_i/V_I$. The coordinates for the $n_i$
points were then determined randomly within the region of volume $v_i$. The
less efficient approach, which was much easier to implement, used a (double)
rejection method similiar to the method described in [9]. In this second
approach, the interval was enclosed in a box and spacetime coordinates were
randomly selected within this box; if the selected point was outside the
interval it was rejected, otherwise, it was kept -- this was the first
rejection. In Minkowski space, this first rejection provides a uniform
distribution of points.

In curved spacetimes, points that fell within the interval faced a second
rejection designed to ensure that the points were distributed uniformly with
respect to the volume form $\Omega ^d$. Each point in the interval was
associated with a random number $w$ selected within the range $0<w<\Omega
_{\max }^d$, where $\Omega _{\max }^d$ is the maximum value of the volume
form within the interval $I$. If $w$ was greater than the value of the
volume form evaluated at the point in question, the point was rejected;
otherwise, it was kept. This process continued until $N$ points were
sprinkled into the interval. The sprinklings in 1+1 dimensions used the
first method; all others used the rejection method. In a few cases the two
methods were compared, and produced completely consistent results. That
these methods produced causal sets that correspond to Poisson sprinklings
were verified, in 1+1 dimensions, by chi-squared tests. In all sprinklings,
random numbers were generated using the subroutine ``ran2'' from [9].

Since the main result of this work comes from comparing the behavior of
small sub-intervals between flat and curved spacetimes, we must determine
the pertinent range of sub-interval sizes. This range can be determined from
sprinklings into Minkowski space. Figure 1 shows the results for random
sprinklings of points into intervals of 2-, 3-, and 4-dimensional Minkowski
space. Both the Myrheim-Meyer ($d_{MM}$) and midpoint-scaling ($d_{mid}$)
dimensions were calculated for every closed sub-interval of size $n_i\geq 3$%
. The average value of $d$ was calculated for sub-intervals of a given size.
To decrease the statistical fluctuations, each curve in the figure
represents an average of 15 different sprinklings.

While there are a number of interesting features in this figure, two things
are most relevant to this study. First, we can see that for the
midpoint-scaling dimension the three different Minkowski spaces are
effectively indistinguishable for sub-intervals smaller than $n_i=10$.
Therefore, since all three Minkowski results agree within this size region,
any curved spacetime that behaves like one of these three should also be in
agreement in this region. This fact sets the lower limit for the pertinent
range of comparison with curved spacetimes at $n_i=3$. Second, the general
trends displayed by these curves are typical for all of the results. The
curves for both $d_{MM}$ and $d_{mid}$ rise steeply producing a ``shoulder''
beyond which the curves level off. The locations of the shoulder are clearly
different for the three different spacetimes; therefore, the degree to which
the analogous results for the curved spacetimes match these flat-spacetime
results around this shoulder will be used to determine the local Minkowski
dimension. The broadest shoulder occurs for $d=4$ for which a value of $%
n_i=100$ is sufficient to incorporate. Therefore, a good size range for
seeking local Minkowski behavior for the curved spacetimes studied here is $%
3\leq n_i\leq 100$. I will call this size range the {\it local Minkowski
region}.

The quantitative measure of how well the results for a curved spacetime
matches those of a particular flat spacetime is a relative goodness-of-fit
test using a chi-squared statistic that compares values of $d$ for
sub-intervals of the same size within the local Minkowski region. This
relative measure requires knowledge of how well the different flat-spacetime
results fit each other according to this method. The statistic is calculated
as 
\begin{equation}
\chi _{a,b}^2=\frac 1B\sum\limits_{j=1}^B\frac{\left( O_{ja}-E_{jb}\right) ^2%
}{E_{jb}},  \label{chi-squared}
\end{equation}
where the subscript ($a,b$) means that $a$-dimensional Minkowski space is
being compared against $b$-dimensional Minkowski space. The quantity $B$ is
the number of bins into which the data was divided (either 22 or 30); this
number depends on the bin size (either 4 or 3) which was chosen such that
each ``expected'' value $E_{jb}$ was greater than 5. The $O_{ja}$ are the
``observed'' values. The results of these calculations for the Myrheim-Meyer
dimension are the following: $\chi _{2,3}^2=0.662$, $\chi _{3,2}^2=1.25$, $%
\chi _{2,4}^2=1.77$, $\chi _{4,2}^2=4.20$, $\chi _{3,4}^2=0.365$, and $\chi
_{4,3}^2=0.457$. For the midpoint dimension we also have the following
results: $\chi _{2,3}^2=0.823$, $\chi _{3,2}^2=1.62$, $\chi _{2,4}^2=2.19$, $%
\chi _{4,2}^2=5.51$, $\chi _{3,4}^2=0.459$, and $\chi _{4,3}^2=0.588$. For
both dimension estimators, the best (smallest) result comes from the
comparison of three-dimensional Minkowski space against 4-dimensional
Minkowski space. Therefore, these values will be used to determine the
relative goodness-of-fit of the results for curved spacetimes, 0.365 for the
Myrheim-Meyer calculations and 0.459 for the midpoint calculations.

\subsection{1+1 dimensions}

Figure 2 shows a uniform sprinkling of 512 points into an interval of a
conformally flat spacetime in 1+1 dimensions with conformal factor $\Omega
^2=(xt)^2$. Both the midpoint and Myrheim-Meyer dimension estimators fail
for the full interval giving values of $2.77$ and $2.65$, respectively. The
conformal factor for this spacetime causes the points to be more spread out
in space which is consistent with the overestimates of the dimension. Figure
3 shows a plot of the average midpoint dimension for sub-intervals of
different size averaged over 15 sprinklings of the spacetime shown in Fig.
2. This curve is compared against the results for Minkowski space. Despite
the fact that the full interval values of the dimension estimators are
closer to 3, the behavior for small sub-intervals clearly follows that of
two-dimensional Minkowki space suggesting a local Minkowski dimension of
two. What appears to be happening here is that the small sub-intervals are,
in fact, behaving like causal sets that are embeddable in two-dimensional
Minkowski space; then, as you look at sub-intervals of larger size the
effects of curvature become more important and the flat-spacetime dimension
estimators become less reliable. (A similar plot using the Mryheim-Meyer
dimension shows identical features).

To quantify this result, a goodness-of-fit test is made, using an equation
very similar to Eq. (\ref{chi-squared}), which compares the average
dimension of sub-intervals in the curved spacetime with those in each
dimension of Minkowski space within the local Minkowski region. The curved
spacetime values are taken as the observed and Minkowski space values as the
expected. These results are then compared to the best chi-squared results
from the mutual comparisons of the different dimensions of Minkowski space.
For the curved spacetime shown in Fig. 2, we obtain relative goodness-of-fit
values $\widetilde{\chi }^2\equiv \chi ^2/\chi _{3,4}^2$ of 
\begin{equation}
\begin{tabular}{ll}
$\widetilde{\chi }_{2D,MM}^2=0.00181,$ & $\widetilde{\chi }%
_{2D,mid}^2=0.00119,$ \\ 
$\widetilde{\chi }_{3D,MM}^2=1.88,$ & $\widetilde{\chi }_{3D,mid}^2=1.84,$
\\ 
$\widetilde{\chi }_{4D,MM}^2=4.95,$ & $\widetilde{\chi }_{4D,mid}^2=4.84,$%
\end{tabular}
\end{equation}
where the notation $\widetilde{\chi }_{2D,MM}^2$ means that the curved
spacetime result was compared against two-dimensional Minkowski space using
the average Mryheim-Meyer dimension values relative to the value of $\chi
_{3,4}^2$ for the Mryheim-Meyer dimension; and correspondingly for the
values labeled with the subscript ``mid.'' As defined, this statistic means
that values of $\widetilde{\chi }^2\gtrsim 1$ represents a poor fit
signifying that the two data sets being compared could certainly be
Minkowski spaces differing in dimension by at least 1. Whereas, values of $%
\widetilde{\chi }^2\ll 1$ indicates a good fit with the spacetime in
question. Clearly, the results displayed in Eq. (10) show that the small
sub-intervals offer an excellent fit to those of two-dimensional Minkowski
space. Furthermore, the fits with three- and four-dimensional Minkowski
space are no better, or much worse, than what can be expected between
Minkowski spaces of different dimensions. The conformally flat spacetime for
which the above results are given represents only one of several 1+1
dimensional spacetimes studied. In all cases, the results are similiar to
those given here.

\subsection{2+1 dimensions}

Figure 4 shows a uniform sprinkling of 512 points into an interval of a
conformally flat spacetime in 2+1 dimensions with conformal factor $\Omega
^2=\left( x^2+y^2\right) /t^6$. This example was chosen because it produced
the worse full interval results for all of the 2+1 dimensional spacetimes
studied. It is easier to see what this interval is like from the
projections. The $y-t$ plane shows that more points are located at larger
values of the $y$ coordinate and smaller values of $t$; the $x-t$ plane
shows similar behavior. The projection onto the $x-y$ plane shows that the
points are more crowded in the middle of the interval. This crowding is due
to the preference for small $t$ where the spatial extent of the region is
centralized.

Figure 5 shows a plot of the average Myrheim-Meyer dimension for
sub-intervals of different size averaged over 15 sprinklings of the
spacetime shown in Fig. 4. This curve is compared against the results for
Minkowski space. For this spacetime, the effects of the curvature become
apparent around $n_i=40$. Nevertheless, the result for the curved spacetime
maintains a good approximation to the flat spacetime result within the
designated locally flat region. To verify that the local Minkowski dimension
of this spacetime should be taken to be three, the relative goodness-of-fit
results are 
\begin{equation}
\begin{tabular}{ll}
$\widetilde{\chi }_{2D,MM}^2=3.03,$ & $\widetilde{\chi }_{2D,mid}^2=3.15,$
\\ 
$\widetilde{\chi }_{3D,MM}^2=0.00807,$ & $\widetilde{\chi }%
_{3D,mid}^2=0.00697,$ \\ 
$\widetilde{\chi }_{4D,MM}^2=1.15,$ & $\widetilde{\chi }_{4D,mid}^2=1.14.$%
\end{tabular}
\end{equation}
Here we see clearly that in the locally flat region this spacetime provides
results that give an excellent fit to the results for three-dimensional
Minkowski space. Several other spacetimes in 2+1 dimensions were studied
giving similar results.

\subsection{3+1 dimensions}

Figure 6 shows projections of a uniform sprinkling of 512 points into an
interval of a conformally flat spacetime in 3+1 dimensions with conformal
factor $\Omega ^2=(x^4+y^4+z^4)/t^6$. Figure 7 is the corresponding plot for
the average dimension per size of sub-interval. As with the other cases, the
figure clearly shows that within the locally flat region the curved
spacetime result gives a much better fit to the Minkowski space having the
same dimension. The relative goodness-of-fit results for this case are 
\begin{equation}
\begin{tabular}{ll}
$\widetilde{\chi }_{2D,MM}^2=9.62,$ & $\widetilde{\chi }_{2D,mid}^2=10.4,$
\\ 
$\widetilde{\chi }_{3D,MM}^2=0.825,$ & $\widetilde{\chi }_{3D,mid}^2=0.919,$
\\ 
$\widetilde{\chi }_{4D,MM}^2=0.0382,$ & $\widetilde{\chi }%
_{4D,mid}^2=0.0252. $%
\end{tabular}
\end{equation}
Several other spacetimes in 3+1 dimensions were studied giving similar
results.

\subsection{A causal set generated by transitive percolation}

So far, all of the causal sets used were guaranteed to be faithful because
they were generated by sprinklings into known manifolds. Having established
the approach, it is instructive to apply this method to a causal set
generated by some other means. Ultimately, there will be a quantum dynamics
for generating causal sets, and it will be these causal sets (or
coarse-grained versions of them) whose manifold dimensions we would like to
estimate. Although a quantum dynamics for causal sets does not yet exist,
there is a classical dynamics, due to Rideout and Sorkin [10], which is
proving to be very useful in helping to determine the extent to which causal
sets can encode physical information. So, this classical dynamics provides
an excellent avenue to illustrate how the suggestions presented here might
be used in a more general case when we cannot be sure that the causal set is
faithfully embeddable.

Perhaps the simplest model within the class of models proposed by Rideout
and Sorkin is the one that they have called transitive percolation. The
procedure followed here for generating a random causal set of $N$ elements
via percolation is as follows: (a) assign labels to the $N$ elements; (b)
impose the partial ordering relation $\prec $ onto each pair of elements
with probability $p$. That is, if $i<k$, the probability that element $%
i\prec $ element $k$ is $p$; and (c) enforce the transitivity requirement on
the set. It has been shown that despite the labeling, causal sets generated
in this way are label invariant in that the probability of getting a
particular causal set is independent of how the elements were initially
labeled [10].

For this example, a causal set with $N=512$ and $p=0.0261$ was generated;
these values produce a causal set that has a Myrheim-Meyer dimension of 2.0
when applied to the full causal set [11]. This causal set, however, is not a
causal set interval. Therefore, the largest sub-interval of this 512-element
causal set was used; this sub-interval contained 298 elements. Figure 8 is a
plot of the average Myrheim-Meyer dimensions for the sub-intervals of the
298-element causal set interval taken from the causal set generated by
transitive percolation, compared against 298-element causal sets generated
by random sprinklings in Minkowski space. What stands out in this figure is
that the percolation curve does not closely follow any of the Minkowski
curves in the local Minkowski region. Therefore, not even the small
sub-intervals of this causal set behave as sub-intervals in Minkowski space
do. This suggests that the percolated causal set should not be embeddable
into any (flat or curved) spacetime.

Looking at the Fig. 8 shows that the percolation curve is a closer fit to
the three-dimensional Minkowski curve than it is to the other curves. The
relative goodness-of-fit test yields 
\begin{equation}
\widetilde{\chi }_{2D,MM}^2=1.82,\quad \widetilde{\chi }_{3D,MM}^2=0.308,%
\quad \widetilde{\chi }_{4D,MM}^2=1.952.
\end{equation}
While these results confirm that the percolation curve is a better fit to
the three-dimensional Minkowski result, we can also see that the $\widetilde{%
\chi }_{3D}^2$ value is nearly two orders-of-magnitude worse than what would
be expected based on our study of the random sprinklings. This fact gives
some quantitative weight to the conclusion reached by studying Fig. 8.

\section{Conclusions}

In this paper, I have suggested a method for estimating the manifold
dimension of a causal set that can be faithfully embedded into curved
spacetimes and tested this method for several conformally flat spacetimes.
The method uses flat-spacetime dimension estimators to search for local
Minkowski behavior within the causal set. This approach can be applied to
any causal set, and works independent of the specific properties of a
particular curved spacetime. Very large causal sets are not required.
Furthermore, this approach is invariant under coarse-graining since both the
Myrheim-Meyer and midpoint-scaling dimensions are invariant under
coarse-graining.

Implementation of this procedure can be summarized as follows: (a) form an
interval of size $N$ of the causal set, larger intervals give better
statistics, but they don't need to be extremely large; (b) average the
Myrheim-Meyer (or midpoint-scaling) dimension for sub-intervals of a given
size; (c) perform ramdom sprinklings of size $N$ or greater in 2-, 3-, and
4-dimensional Minkowski space and determine average dimension values for
their sub-intervals; (d) a comparision of the results for the causal set
being checked against the results for the three Minkowski sprinklings for
small sub-intervals, $3<n_i<100$, should reveal whether or not the causal
set in question displays the local Minkowski behavior that would be required
of causal sets that are faithfully embeddable into physically relevant
spacetimes.

It is worth noting that instead of calculating the dimension values for
closed sub-intervals, open sub-intervals can also be used. However, the
statistical fluctuations are greater for open sub-intervals and this fact
becomes somewhat problematic in 2+1 and especially in 3+1 dimensions. What
now remains is to apply this method to generically curved spacetimes. The
basic principle behind the local Minkowski dimension certainly applies in
the generic case, but the extent to which this behavior can be extracted
from the causal sets is yet unknown. If this approach proves useful in that
case as well, it would be an important step toward the goal of a more
comprehensive manifold test for causal sets.

\section{Acknowledgments}

I'd like to acknowledge the help of Mr. Jason Ruiz for writing the computer
code to conduct the chi-squared tests used to test the faithfulness of some
of the sprinklings. I also wish to thank Dr. Rafael D. Sorkin for providing
information on the percolation dynamics and further motivation to work on
problems in causal set quantum gravity. This work was supported by the
Graduate Research Fund and a Faculty Research Fellowship both from Eastern
Michigan University. Additional support from the National Science Foundation
is gratefully acknowledged.

\section{Figure Captions}

FIG 1. The average value of the Myrheim-Meyer and midpoint dimensions for
sub-intervals of a given size in 1+1, 2+1, and 3+1 dimensional Minkowski
space. The sprinklings in 1+1 and 2+1 dimensions are of 512 points while,
for better statistics, the 3+1 sprinklings were of 1024 points. Each curve
is an average of 15 different sprinklings. To clearly see the behavior of
the small sub-intervals results for sub-interval containing only $n_i\leq
200 $ are shown here. For $n_i>200$ the results for the Myrheim-Meyer and
midpoint dimensions, for the 2+1 and 3+1 sprinklings, also merge to the
appropriate interger values.

FIG 2. A uniform sprinkling of 512 points into an interval of a conformally
flat spacetime in 1+1 dimensions. The conformal factor is shown in the
figure.

FIG 3. Comparison of the average value of the midpoint-scaling dimension for
sub-intervals of a given size for the set of points shown in Fig. 2 (2D
curved) against the similar results for 2-, 3-, and 4-dimensional Minkowski
space. The results for small sub-intervals suggests a local Minkowski
dimension of two.

FIG 4{\bf .} A uniform sprinkling of 512 points into an interval of a
conformally flat spacetime in 2+1 dimensions with conformal factor $\Omega
^2=(x^4+y^4)/t^6$. The figure also shows projections of the points onto the $%
x-t$, $y-t$, and $x-y$ planes.

FIG 5. Comparison of the average value of the Mryheim-Meyer dimension for
sub-intervals of the set shown in Fig. 4 (3D curved) against similar results
for Minkowski space. The results for small sub-intervals suggests a local
Minkowski dimension of three.

FIG 6{\bf .} Projections of a uniform sprinkling of 512 points into an
interval of a conformally flat spacetime in 3+1 dimensions with conformal
factor $\Omega ^2=(x^4+y^4+z^4)/t^6$. Panel (a) shows the projection onto
the $x-t$ plane, panel (b) shows the projection onto the $y-t$ plane, panel
(c) shows the projection onto the $z-t$ plane, and panel (d) shows the
projection onto the $x-y$ plane.

FIG 7. Comparison of the average value of the midpoint-scaling dimension for
sub-intervals of the set represented in Fig. 6 (4D curved) against similar
results for Minkowski space. The results for small sub-intervals suggests a
local Minkowski dimension of four.

FIG 8. Comparison of the average value of the Mryheim-Meyer dimension for
the largest sub-interval of a 512-element causal set generated by transitive
percolation. The results for small sub-intervals do not match any of the
three Minkowski space results suggesting that this percolated causal set is
not faithfully embeddable into any spacetime manifold.


\section{References}

\begin{references}
\bibitem{1}  L. Bombelli, J. Lee, D. Meyer, and R. D. Sorkin, Phys. Rev.
Lett. {\bf 59}, 521-24 (1987); Phys. Rev. Lett. {\bf 60}, 656 (1988).

\bibitem{2}  D. D. Reid, Can. J. Phys. {\bf 79}, 1-16 (2001).

\bibitem{3}  J. Myrheim, CERN preprint TH-2538 (1978).

\bibitem{4}  D. Meyer, ``{\it The Dimension of Causal Sets},'' Ph.D. thesis,
M.I.T. (1988).

\bibitem{5}  L. Bombelli, ``{\it Space-time as a Causal Set},'' Ph.D.
thesis, Syracuse University (1987).

\bibitem{6}  M. H. DeGroot, ``{\it Probability and Statistics},''
Addison-Wesley (Reading, 1975).

\bibitem{7}  G. Brightwell and R. Gregory, Phys. Rev. Lett. {\bf 66},
260-263 (1991).

\bibitem{8}  R. D. Sorkin. {\it In} ``{\it General Relativity and
Gravitational Physics. Proceedings of the Ninth Italian Conference on
General Relativity and Gravitational Physics},'' September 1990. {\it Edited
by} R. Cianci, R. de Ritis, M. Francaviglia, G. Marmo, C. Rubano, and P.
Scudellaro. World Scientific (Singapore, 1991), pp. 68-90.

\bibitem{9}  W. H. Press, S. A. Teukolsky, W. T. Vetterling, and B. P.
Flannery, ``{\it Numerical Recipes in Fortran: The Art of Scientific
Computing},'' 2nd. ed., Cambridge University Press (Cambridge, 1992).

\bibitem{10}  D. P. Rideout and R. D. Sorkin, Phys. Rev. D {\bf 61}, 024002
(2000).

\bibitem{11}  R. D. Sorkin, private communication.
\end{references}
\end{document}